\documentclass[12pt,preprint]{aastex}
\shorttitle{MACHO Project LMC Cepheid Eclipsing Binaries}
\shortauthors{Alcock et al.}
\begin{document}

\title{The MACHO Project LMC Variable Star Inventory:\\  
       XII. Three Cepheid Variables in Eclipsing Binaries}

\author{
      C.~Alcock\altaffilmark{1,2,3},   
    R.A.~Allsman\altaffilmark{4},      
    D.R.~Alves\altaffilmark{5},        
    A.C.~Becker\altaffilmark{6,7},     
    D.P.~Bennett\altaffilmark{8},      
    K.H.~Cook\altaffilmark{1,2},       
    A.J.~Drake\altaffilmark{1,9},      
    K.C.~Freeman\altaffilmark{9},      
      K.~Griest\altaffilmark{2,10},    
    S.L.~Hawley\altaffilmark{6},       
      S.~Keller\altaffilmark{1},	
    M.J.~Lehner\altaffilmark{11},      
      D.~Lepischak\altaffilmark{12},   
    S.L.~Marshall\altaffilmark{1,2},   
      D.~Minniti\altaffilmark{13},     
    C.A.~Nelson\altaffilmark{1,14},    
    B.A.~Peterson\altaffilmark{9},     
      P.~Popowski\altaffilmark{15},     
    M.R.~Pratt\altaffilmark{6,16},        
    P.J.~Quinn\altaffilmark{17},       
    A.W.~Rodgers\altaffilmark{9,18},   
    N.~Suntzeff\altaffilmark{19},      
      W.~Sutherland\altaffilmark{20},  
      T.~Vandehei\altaffilmark{2,10},   
    D.L.~Welch\altaffilmark{12}\\      
    {\bf (The MACHO Collaboration) }
}

\altaffiltext{1}{Lawrence Livermore National Laboratory, Livermore, CA 94550\\
    Email: {\tt kcook, adrake, keller19, stuart, cnelson@igpp.llnl.gov}}

\altaffiltext{2}{Center for Particle Astrophysics, University of California, Berkeley, CA 94720}

\altaffiltext{3}{Department of Physics and Astronomy, University of Pennsylvania,\\
209 South 33rd St., Philadelphia, PA 19104-6396\\
  Email: {\tt alcock@hep.upenn.edu}}

\altaffiltext{4}{Supercomputing Facility, Australian National University,
    Canberra, ACT 0200, Australia \\
    Email: {\tt Robyn.Allsman@anu.edu.au}}

\altaffiltext{5}{Columbia Astrophysics Laboratory, 550 W. 120th St.,
    NY, NY 10027\\
    Email: {\tt alves@astro.columbia.edu}}

\altaffiltext{6}{Departments of Astronomy and Physics,
    University of Washington, Seattle, WA 98195\\
    Email: {\tt hawley, pratt@astro.washington.edu}}

\altaffiltext{7}{Bell Laboratories, Lucent Technologies, 
    600 Mountain Avenue, Murray Hill, NJ 07974\\
    Email: {\tt acbecker@physics.bell-labs.com}}

\altaffiltext{8}{Department of Physics, University of Notre Dame, IN 46556\\
    Email: {\tt bennett@bustard.phys.nd.edu}}

\altaffiltext{9}{Research School of Astronomy and Astrophysics, 
        Canberra, Weston Creek, ACT 2611, Australia\\
 Email: {\tt kcf, peterson@mso.anu.edu.au}}

\altaffiltext{10}{Department of Physics, University of California,
    San Diego, CA 92093\\
    Email: {\tt kgriest@ucsd.edu, vandehei@astrophys.ucsd.edu }}

\altaffiltext{11}{Department of Physics, University of Sheffield, Sheffield S3 7RH, UK\\
    Email: {\tt m.lehner@sheffield.ac.uk}}

\altaffiltext{12}{McMaster University, Hamilton, Ontario Canada L8S 4M1\\
    Email: {\tt lepischak, welch@physics.mcmaster.ca}}

\altaffiltext{13}{Depto. de Astronomia, P. Universidad Catolica, Casilla 104, 
        Santiago 22, Chile\\
Email: {\tt dante@astro.puc.cl}}

\altaffiltext{14}{Department of Physics, University of California, Berkeley,
        CA 94720}

\altaffiltext{15}{Max-Planck-Institut f\"{u}r Astrophysik, Karl-Schwarzschild-Str.\ 1, Postfach
1317, 85741 Garching bei M\"{u}nchen, Germany.\\
Email: {\tt popowski@mpa-garching.mpg.de}}

\altaffiltext{16}{Center for Space Research, Massachusetts Institute of Technology, Cambridge, MA 02139}

\altaffiltext{17}{European Southern Observatory, Karl Schwarzchild Str.\ 2, 
        D-8574 8 G\"{a}rching bel M\"{u}nchen, Germany\\
Email: {\tt pjq@eso.org}}

\altaffiltext{18}{Deceased.}

\altaffiltext{19}{National Optical Astronomy Observatories,Cerro Tololo Inter-American Observatory, 
Casilla 603, La Serena, Chile\\
	Email:  {\tt nsuntzeff@noao.edu }}

\altaffiltext{20}{Department of Physics, University of Oxford,
    Oxford OX1 3RH, U.K.\\
    Email: {\tt w.sutherland@physics.ox.ac.uk}}


\begin{abstract}

We present a method for solving the lightcurve of an eclipsing binary
system which contains a Cepheid variable as one of its components as
well as the solutions for three eclipsing Cepheids in the Large
Magellanic Cloud (LMC).  A geometric model is constructed in which the
component stars are assumed to be spherical and on circular orbits.
The emergent system flux is computed as a function of time, with the
intrinsic variations in temperature and radius of the Cepheid treated
self-consistently.  Fitting the adopted model to photometric
observations, incorporating data from multiple bandpasses, yields a
single parameter set best describing the system.  This method is
applied to three eclipsing Cepheid systems from the MACHO Project LMC
database: MACHO ID's 6.6454.5, 78.6338.24 and 81.8997.87.  A best-fit
value is obtained for each system's orbital period and inclination and
for the relative radius, color and limb-darkening coefficients of each
star.  Pulsation periods and parameterizations of the intrinsic color
variations of the Cepheids are also obtained and the amplitude of the
radial pulsation of each Cepheid is measured directly.  The system
6.6454.5 is found to contain a 4.97-day Cepheid, which cannot be
definitely classified as Type I or Type II, with an unexpectedly
brighter companion.  The system 78.6338.24 consists of a 17.7-day, W
Vir Class Type II Cepheid with a smaller, dimmer companion.  The
system 81.8997.87 contains an intermediate-mass, 2.03-day overtone
Cepheid with a dimmer, red giant secondary.

\end{abstract}



\section{INTRODUCTION}

Large scale microlensing surveys have provided unprecedented resources
for variable star research.  Their long time baseline and stable,
accurate photometry are ideal for the detection and analysis of such
objects and the large number of systems observed increases the
probability of finding astrophysically-interesting objects that either
have escaped detection or do not exist in our own galaxy.

Very few regularly-pulsating stars are known to belong to eclipsing
systems.  One of the best studied is AB Cas, an Algol-type binary
system, which contains a $\delta$ Scuti type primary \citep{abcas}.
\citet{R00} list 6 additional $\delta$ Scutis which are members of
eclipsing binaries.  The lone Galactic candidate for an eclipsing
binary containing a Cepheid variable was BM Cas \citep{Thiessen} but
further study revealed that the variable was unlikely to be a Cepheid
\citep{fernev}.  The astrophysical benefits of a Cepheid variable in
an eclipsing system could be considerable.  If the system is
double-lined, a determination of the Cepheid's luminosity and mass can
be made that is not only more accurate than existing measurements but
also truly independent of the intervening steps in the distance
ladder.  Such a system would provide the most direct measurement of
the mass of a Cepheid and would offer an independent calibration of
the period-luminosity relation.

Here we present the results of lightcurve analyses of three eclipsing
Cepheid systems in the MACHO project Large Magellanic Cloud (LMC)
database: MACHO ID's 6.6454.5, 78.6338.24 and 81.8997.87.  In Sec. 2
we describe the sources of the photometric observations.  Sec. 3
describes the model used to generate the lightcurve of an eclipsing
Cepheid system and Sec. 4 describes the inverse problem of computing
the parameters from an observed lightcurve.  In Sec. 5 we present the
results obtained for the three systems.  Finally, Sec. 6 summarizes
the analysis, describes work in progress and suggests future avenues
for research.


\section{OBSERVATIONS}

Observations from several different sources are used in the analysis
presented here but the majority are from the MACHO Project database.
The MACHO observations were made with the refurbished 1.27m Great
Melbourne Telescope at Mount Stromlo, near Canberra, ACT, Australia.
It has been equipped with a prime focus reimager-corrector with an
integral dichroic beamsplitter to give a 0.5 sq. deg field of view in
two passbands simultaneously: a 450-590 nm MACHO $V$ filter and a
590-780 nm MACHO $R$ filter.  These are each sampled with a 2$\times$2
array of 2048$\times$2048 Loral CCDs which are read out concurrently
via two amplifiers per CCD in about 70 seconds.  The resulting images
cover 0.5 square degrees with 0.63 arcsec pixels \citep{al95}.  Data
reduction is performed automatically by Sodophot, a derivative of
DoPhot \citep{dophot}.  MACHO photometry is then transformed into
Cousins $V$ and $R$ bands for further interpretation \citep{al99}. The
eclipsing, pulsating nature of systems 6.6454.5 and 78.6338.24 was
identified through visual inspection of their lightcurves.

The eclipsing nature of 81.8997.87 was first reported by the Optical
Gravitational Lensing Experiment (OGLE) project \citep{ogle} (OGLE ID
LMC\_SC16 119952) and the lightcurves for this system are a
combination of observations in the OGLE and MACHO databases.  OGLE
observations were taken on the 1.3 m Warsaw telescope at Las Campanas
Observatory, Chile operated by the Carnegie Institution of Washington.
Photometry is in the standard $BVI$ bands with the majority of the
observations in the $I$ band \citep{ogle}.  $V$ and $I$ measurements
for 6.6454.5 (LMC\_SC21 40876) have also been used here.  The OGLE
observations of the April 2000 eclipse of this system were graciously
provided by Andrzej Udalski in advance of publication.

For the 1999 eclipse, observations of 6.6454.5 were taken by Nick
Suntzeff from March 16 to April 21 on the 0.9m telescope at the Cerro
Tololo Inter-American Observatory in Chile.  The telescope is a
0.9-meter Ritchey-Chretien Cassegrain telescope with a dedicated
2048$\times$2048 CCD detector.  Observations were in standard $BVI$
and were reduced by DAOPHOT, ALLSTAR and ALLFRAME.

A complete listing of the observations, from all sources, used in this
paper is available at
\url{http://www.macho.mcmaster.ca/Data/EclCep/ecl\_cep.htm}.  For
guidance regarding form and content a partial listing is given in
Tables \ref{tab-6.6454.5v} - \ref{tab-81.8997.87i}.


\section{MODEL}

To analyse the lightcurve of eclipsing Cepheid systems, a geometric
model based on classical equations \citep[for instance]{binne} was
developed.  In this model the stars are assumed to be traveling on
circular orbits.  This isn't the most general case but may be
appropriate here.  It is found that two out of the three systems
studied here have their primary and secondary minima separated by
almost exactly one half period, consistent with a circular orbit.  The
stars themselves are treated as circular disks with radius ($R$) and
central surface brightness ($J_\circ$).  A linear limb-darkening law
of the form
\begin{equation}  
J = J_\circ ( 1 - x_{\lambda} + x_{\lambda}\cos(\gamma)), \label{eq:ld}
\end{equation}  
is adopted where $x_\lambda$ is the limb-darkening coefficient and
$\gamma$ is the angle between the surface normal at that point and the
line of sight.  Integrating over $\gamma$ yields the light received from a non-eclipsed star as being proportional to
\begin{equation}
L = \pi R^2 J_\circ \left(1 - \frac{x_\lambda}{3}\right) \label{eq:lcal}
\end{equation}
As absolute dimensions of the stars cannot be extracted from the
lightcurve alone, the above parameters all represent ratios of
observed quantities: the radii are measured in units of the orbital
separation of the two stars, the surface brightnesses are normalized
such that the mean light received outside of eclipse $(L_1 + L_2)$ is
1.0.

In many programs the values of $J_\circ$ are modelled directly with a
different value for each bandpass.  As we have simultaneous MACHO $V$
and $R$ observations for all systems at all phases, we fit the colour,
specifically the ratio $J_V/J_R$ for each star, rather than the
surface brightness.  From this surface brightness ratio the usual
$V-R$ colour index can be calculated.  If I band observations are
available, as they are for two of the three systems, the flux ratio
$J_V/J_I$ must also be defined, but is not used as a fitted parameter.
Instead the $V-I$ colour index is computed from the $V-R$ colour using
a linear relationship derived from the $VRI$ observations of
\citet{cousins} of a sample of southern stars.  From the colour index,
$J_V/J_I$ can be calculated.  To convert the surface brightness ratios
into values for the individual surface brightnesses we employ the
linear relationship between V surface brightness ($J_V$) and the $V-R$
colour index identified by \citet{BE} (hereafter B-E).  This
relationship holds for stars of all spectral types and for all
pulsational phases of the Cepheid.  Unfortunately, it is not certain
that the B-E relation applies to metal-poor Type II Cepheids and this
is a potential source of systematic error in our results.  However,
not all Type II Cepheids are metal poor: one of the few Type II
Cepheids known to be in spectroscopic binaries, the 2.4-day Type II
Cepheid AU Peg \citep{1984AJ.....89..119H}, has an [Fe/H] = +0.1.  The
B-E relation applies to Johnson filter definitions so the
transformations of \citet{Bell} were used to convert to forms
applicable to our Cousins bandpasses.  With values of $J_V$ the other
surface brightnesses follow from the ratios.
    
The intrinsic variability of the Cepheids is handled by taking $R$ and
$J_V/J_R$ as the minimum radius and the mean colour, respectively.
Variability is then added to these values in some functional form.
The colour variation is parameterized by a third-order Fourier series:
\begin{equation}  
\Delta\left(\frac{J_V}{J_R}\right) = \sum_{k=1}^3 A_k \cos(k \omega t) + B_k \sin(k \omega t) \label{eq:fou1}
\end{equation}  
The radial variation is modeled by the expression:
\begin{equation}
\Delta R = a \left|\sin\left(\frac{1}{2} \omega (t-t_\circ)\right)\right|
\end{equation}
This produces a curve which, despite its discontinuity at $\Delta R =
0$, reproduces the broad features of the radial variation curve.

During eclipses, the decrease in the amount of light received from the
system is calculated analytically, based on the radii of two
overlapping discs and geometrical considerations \citep{binne}.  The
apparent separation of the centres of the two stars is given by
\begin{equation}  
\delta^2 = \cos^2i + \sin^2i~\sin^2\theta
\end{equation}  
Whether the system is in eclipse or not is determined by comparing
this distance to the instantaneous radii of the stars.  Within
eclipse, the area obscured by the eclipsing star is given by
\begin{eqnarray} 
Area&=&\frac{1}{2}R_1^2 \left(2 \alpha_1-\sin(2 \alpha_1) \right) + \frac{1}{2} R_2^2 \left( 2
\alpha_2-\sin(2 \alpha_2) \right) \label{eq:a1}\\ 
\cos(\alpha_1)&=&\frac{R_1^2-R_2^2+ \delta^2}{2R_1 \delta}\label{eq:a2}\\
\cos(\alpha_2)&=&\frac{R_2^2-R_1^2+ \delta^2}{2R_2 \delta}\label{eq:a3}
\end{eqnarray}
Multiplying the area by the surface brightness of the eclipsed star
gives the decrease in flux.

The effects of limb-darkening within the eclipses are included by
dividing the stellar disk into 100 concentric rings in a manner
similar to \citet{nelsondavis}.  A surface brightness is then assigned
to each ring based on equation \ref{eq:ld}.  The total light received
from each star can then be calculated by computing the contribution
from each ring and integrating over all of the rings.  Within eclipses
the integration must be performed within suitable limits based on the
amount of each stellar surface that is visible.  In practice, it is
faster to compute the normal fluxes from equation \ref{eq:lcal} and
then calculate the amount of flux lost based on the eclipse geometry.
This is done by using equations \ref{eq:a1}-\ref{eq:a3} repeatedly for
eclipsed stars of varying radii, which correspond to the rings into
which the stellar disk is divided, and then summing over the total
number of rings.  So the flux lost would be given by
\begin{equation}
\sum_{i=0}^{100}J_i \left( Area(r_i)-Area(r_{i-1}) \right)
\end{equation}
where $r_0$ to $r_{100}$ are the radii of individual rings ranging
from 0 to the radius of the eclipsed star ($R_1$), $Area(r_i)$ is the
area of a disk of radius $r_i$ covered by the eclipsing star, given by
equations \ref{eq:a1}-\ref{eq:a3}, and $J_i$ is the surface brightness
of that ring given by equation \ref{eq:ld}.  This flux lost is then
subtracted from the flux given by equation \ref{eq:lcal} to give the
eclipsed star's contribution to the sytem flux.  Integration over the
eclipsed region would be more accurate in principle, but poses the
practical problem of being more difficult to implement.  The limits of
integration are particularly difficult to express for a star of
varying radius as $R_1$ (as well as $J_\circ$ in equation \ref{eq:ld})
will be a function of time.  Summation is more straightforward to
implement and the loss of accuracy in the final flux value is, with
the number of annuli being used, on the order of one part in 1000, as
determined by comparison of trial summations over a full, uneclipsed
disk with the result from the analytic expression (equation
\ref{eq:lcal}).


\section{FITTING PROCEDURE}

Prior to the fitting of the adopted model to the observed lightcurves,
a preliminary analysis of each lightcurve was performed.  This was
done to produce a set of initial parameters as a starting point for
the fitting process.  The orbital period can easily be estimated by
measuring the average length of time between eclipses.  Then, a coarse
removal of the pulsation contribution to the lightcurve is achieved by
phasing the data to the orbital period and fitting hyperbolas to both
the primary and secondary eclipses.  The out-of-eclipse flux is
assumed to be 1.0.  The result is an averaged lightcurve which gives
clearly defined estimates for the average minima of both eclipses and
average phase of external contact.  Values for the inclination, radii
and relative luminosities were obtained using an algorithm from
\citet{riazi}: from the values for the system light, in two filters,
at the eclipse minima, this algorithm gives the relative luminosities
of the two stars as well as the ratio of radii.  From these, and the
approximate phase angle at external contact, the inclination and
relative radii can be computed.  These values are then improved by
performing a fit to minimize $\chi^2$ relative to a simplified model.
Once the boundaries of the eclipses have been defined the parameters
controlling the Cepheid variability can be estimated by examining only
the out-of-eclipse data.  The pulsation period is found using a
routine {\tt period} from \citet{nr} on the out-of-eclipse data
points.  This routine uses an algorithm by \citet{lomb} to compute the
power present at various frequencies in unevenly spaced data. Once the
pulsation period is known, the data outside of eclipse can be phased
to it and a Fourier series fit to the flux variation.  The initial
radial variation is simply set to be one quarter of the radius of the
variable star and the offset is set to be one quarter of a period
later than the time zeropoint for the lightcurve.  The parameter set
resulting from the above procedures will not accurately describe the
lightcurve but it is usually sufficient to produce a lightcurve close
enough to that observed to allow fitting to proceed and to be in the
same valley in $\chi^2$ parameter space as the global minimum.

The data are then subjected to a $\chi^2$ minimization by means of a
standard Levenberg-Marquandt method from \citet{nr}.  This procedure
alternates between two complementary methods for finding a minimum
$\chi^2$: the steepest descent method far from the minimum, switching
smoothly to the inverse Hessian method as the minimum is approached.
This approach combines the advantages of both methods: rapid progress
is made towards a minimum and, once located, accurate determination of
the best-fit parameters is achieved.  The drawback is the possibility
of termination within a local minimum rather than the true minimum.
This possibility can be reduced through reasonable selection of the
initial values for the fitting process and examination of the final
parameter set.  An array of flags allows each parameter to be fit or
held fixed.  The data from both filters (or all three filters, when
available) are fit simultaneously, allowing parameters such as the
inclination and the radii, which correspond to physical dimensions of
the system (and so should not vary from filter to filter) to be
determined from all of the data.

The limb-darkening coefficients prove to be the parameters most
difficult to fit and have to be treated separately.  A fit is first
performed with all limb-darkening coefficients held fixed at their
initial values, usually 0.5, to produce improved values for the other
parameters.  Another fit is then performed with the limb-darkening
coefficients allowed to vary to produce optimum values for all
parameters.  The $x$ will not vary far from 0.5, typically staying
within 0.4 - 0.7, the range expected for $x$ of most main-sequence and
giant stars.  The relative movement of the $x_\lambda$ values can then
be examined to see if it matches expectations based on the relative
temperatures and surface gravities of the two stars.  In practice the
uncertainties on the limb-darkening coefficients prove to be too large
for meaningful analysis.  More ``in eclipse'' data would improve the
constraints on these coefficients.
 
To distinguish between primary-variable and secondary-variable
configurations, we solved for the system parameters with either
configuration in turn.  The set with the lower $\chi^2$ was adopted.

We also investigated the possibility of sources of flux in each system
in addition to the two eclipsing components.  This involved assuming a
third light source, of various brightnesss and refitting the system.
In each of the three systems this produced no significant improvement
in the fit.


\section{RESULTS AND DISCUSSION}

The best fit parameters for each of the systems are shown in Tables
\ref{tab:6param}-\ref{tab:81param}.  The orbital and pulsational
periods are both in days and the inclination is given in degrees.  The
radii (R) and amplitude of the radial variation ($\Delta R_{amp}$) are
relative to the orbital separation of the two stars.  $\Delta
R_{shift}$, which measures the shift of the radial change relative to
the temperature change, has units of days.  Only the ratio of surface
brightnesses $J_V/J_R$ is tabulated as it was the only one that was
fit directly and the other surface brightnesses are computed from it
as outlined above.  The limb-darkening coefficients, $x_\lambda$ are
as defined in equation \ref{eq:ld}.  Once the relative surface
brightness and radius of each star has been determined they can be
combined with the mean system magnitude to compute the magnitude of
each star in all filters.  Also computed for each of the three
Cepheids is the value of $W_R = R - 4.0(V-R)$, an index which corrects
for most of the effects of reddening and effective temperature
differences.



Figures \ref{fig:6vp}-\ref{fig:81vp} show the primary eclipses for
each system along with the best fit lightcurve.  Figure \ref{fig:pl}
shows a period-$W_R$ diagram (P-L diagram) for MACHO Cepheids and the
locations of the three Cepheids studied here.  This diagram is
essentially free of reddening and allows us to classify the three
Cepheids under study based on their relation to other LMC Cepheids
without an explicit correction for extinction.  The uncertainty ranges
in the magnitude values in Tables \ref{tab:6param}-\ref{tab:81param}
and Figure \ref{fig:pl} were estimated from the range of possible
component magnitudes based on the uncertainties in the best fit
parameters.  These are statistical uncertainties and likely
underestimate the true uncertainties.

The colors and magnitudes of each star allow some general comments on
the evolutionary state of each system if we assume each star follows a
standard, single-star evolutionary history.  This assumption is not
unwarranted here given the large orbital periods of the systems.
First, a crude correction for extinction must be made. To account for
foreground reddening values of $E(B-V)$ are adopted from the map of
Galactic foreground color excess toward the LMC published by
\citet{sch91}.  This yields values of $E(B-V)$ for the three systems
as follows: 0.08 for 6.6454.5, 0.08 for 78.6338.24 and 0.10 for
81.8997.87.  These allow us to determine values for $A_V$ and $A_R$
when combined with the standard value of the ratio of total to
selective extinction, $R_V = A_V/E(B-V) = 3.1$ \citep{cousins}, and
$A_R/A_V = 0.77$ for our Cousins $R$ and $V$ from the interstellar
extinction relations of \citet{cardelli}.

This procedure is clearly sufficient for the system 78.6338.24, producing
$V_\circ = 16.20 \pm 0.03$ which agrees with the period-$M_v$ relation
of \citet{type2} for type II stars with log $P > 1.1$.  Applying the
correction described above to the system 81.8997.87 fails to produce V
and R magnitudes for the Cepheid that are consistent with those
expected of an overtone Cepheid of its period.  Given the system's
proximity to the 30 Doradus star-forming region it is not unreasonable
to expect substantial extinction along this line of sight within the
LMC and the closest Cepheid on the same plate, the overtone
81.8997.128, also appears well below the overtone band in the P-L
diagram prior to applying a reddening correction.  The
period-magnitude relation for overtone Cepheids of \citet{baraffe}
gives $V_\circ = 15.86$ for the Cepheid in 81.8997.87, implying $A_v =
1.31$, a total value of $E(B-V) = 0.42$ and $R_\circ = 15.56$.  Both
$V_\circ$ and $R_\circ$ are consistent with the overtone bands in the
period-V magnitude and period-R magnitude plots for MACHO LMC
Cepheids.  A total value of $E(B-V) = 0.41$ was obtained by
\citet{demarchi} for selected regions of 30 Doradus.

A further correction may also be necessary for the system 6.6454.5,
however, the ambiguity in the classification of the Cepheid (see
below) precludes definitively comparing its properties to those
expected from a period-luminosity relation as was done with the other
two systems.  In light of this we adopt only the correction for
foreground reddening.

After correction for reddening the values of $V$ and $V-R$ are
converted to the $L - T_{eff}$ plane by assuming $\mu_{LMC} = 18.5$
mag and using
\begin{equation}
log(T_{eff}) = 4.199 - \sqrt{0.08369+0.3493(V-R)}
\end{equation}
a transformation of the \citet{cwc} semi-empirical calibration.
Figures \ref{fig:iso1} and \ref{fig:iso2} show theoretical isochrones
representative of the metallicities of Type I and Type II stellar
populations: Y=0.25, Z=0.008 isochrones from \citet{bertelli} in
Figure \ref{fig:iso1} and Y=0.230, Z=0.0004 isochrones from
\citet{fagotto} in Figure \ref{fig:iso2}, along with the properties of
the three systems.  The error bars are computed exclusively from the
errors in the magnitudes and colors given in Tables
\ref{tab:6param}-\ref{tab:81param}.

Based on the properties tabulated in Table \ref{tab:6param} the system
6.6454.5 is found to contain a Cepheid as the secondary with a
brighter, bluer primary.  The nature of the Cepheid is unclear as its
location in the P-L diagram (Figure \ref{fig:pl}) places it between
the Type I and Type II bands, inconsistent with both classifications.
The large amplitude of the radial variation, 0.323 $\pm$ 0.008 of the
Cepheid's minimum radius is more consistent with a Type II
classification.  Fundamental-mode Type I Cepheids have typical radial
variations of of 10\% or less \citep{armstrong01} while Type II
Cepheids show larger radial excursions in the range of 30-50\% of the
minimum radius \citep{leb92}.

If the Cepheid is assumed to be a Type II Cepheid it is either making
an excursion from the AGB or moving off the HB to the AGB (a less
likely scenario given its period).  The companion, displayed in Figure
\ref{fig:iso2}, which also appears to be considerably evolved, is too
luminous to fit either of these scenarios.  It is possible that this
system is not in the LMC but is instead a foreground object.  A
reduction in the assumed distance to the system to 17.8 kpc is
necessary to shift the Cepheid's properties to fit a post-HB
evolutionary state.  By contrast, if the system is compared to Type I
isochrones (Figure \ref{fig:iso1}) the Cepheid's location is
consistent with that expected but the companion appears to be too blue
to fit the isochrones.  If the Cepheid is indeed Type I, an additional
reddening correction would need to be applied to make its location in
Figure \ref{fig:pl} consistent with the fundamental, Type I band.
This would imply an even higher effective temperature and luminosity
for the companion.

The status of 78.6338.24 is less ambiguous. It consists of a Type II
Cepheid secondary with a hotter, but somewhat dimmer primary.  With a
pulsational period in excess of 17.5 days the Cepheid would classified
as a W Vir type, which is consistent with its large radius relative to
its companion and with its location in the P-L diagram (Figure
\ref{fig:pl}).  This system presented several challenges to modelling.
The pulsation period of the Cepheid was found to not be constant over
the duration of the observations.  Despite being very small in
magnitude (less than 1\% of the period) this drift in period produced
a substantial decrease in the quality of the fit.  It was corrected to
some degree by assuming a pulsational angular frequency that was a
slowly varying fuction of time, parameterized by:
\begin{equation}
\omega = \omega_\circ + A_1Bt + A_2(Bt)^2 \label{eq:freq}
\end{equation}
where the $A_i$ are parameters to be determined by fitting and B is a
constant, set by trial and error, to ensure that the $A_i$ are of the
same order as the other parameters.  Inspection of the complete set of
residuals after fitting revealed indications of non-sphericity in one
(or both) of the system components.  Both the asphericity and the
``period drift'' are consistent with the large radius and tenuous
outer envelope of a W Vir star.  Their locations in Figure
\ref{fig:iso2} show both components to be well-evolved, post-HB or
post-AGB objects.

The binary 81.8997.87 is distinct from the other two systems in
several ways.  Its variable is an intermediate-mass Cepheid pulsating
in the first-overtone mode.  Furthermore, the Cepheid is the primary
with a considerably cooler, dimmer companion.  The amplitude of radial
pulsations is low, 0.060 $\pm$ 0.006 of the minimum Cepheid radius, as
expected for an overtone Cepheid.  \citet{gie82} provides an explicit
determination of the radial displacement for the galactic overtone
Cepheid SU Cas.  The radial amplitude of that star is 0.026 of the
mean Cepheid radius, similar to the value we find for 81.8997.87 (note
that workers tend to avoid Baade-Wesslink analyses of overtone
Cepheids due to the small dynamic range of the observables).  Table 5a
of \citet{1978A&A....62...75P} lists physical properties derived from
fitting model atmospheres to the continuum colors of a sample of
Galactic Cepheids.  The values of $\Delta R / \langle R \rangle$
obtained for the six overtones in the sample range from 0.042 to
0.080.  Once photometric contamination from the companion is removed
the Cepheid's lightcurve shape, parameterized by the Fourier ratio
$R_{21}$, is consistent with those of other MACHO overtone Cepheids of
similar period.  Its location in the P-L diagram ({Figure
\ref{fig:pl}) further reinforces this view.  The companion's
properties suggest a RGB star, possibly K class or later, however this
combination of binary components is in poor agreement with current
models of single star evolution.  This is reflected in the
disagreement with the isochrones seen in Figure \ref{fig:iso1}.
Indeed any system consisting only of these two stars may be
evolutionarily inconsistent.

The observational coverage of this system is far from ideal.  The
800-day orbital period limits the number of primary eclipses in the
MACHO database to only 3.  Furthermore the secondary eclipses appear
to be non-existent.  This can be accounted for by assuming a very low
surface brightness for the secondary which will produce a very shallow
secondary eclipse that could be dwarfed by the Cepheid variability.
The absence of significant secondary eclipses could be a consequence
of poor observational coverage.  The near 2-day pulsational period
(2.03 days) combined with MACHO's single-point per night coverage
results in repeated sampling of the same two pulsation phases during
an individual eclipse.  These factors, poor coverage and the
ill-defined secondary eclipses, produce large uncertainties in the fit
parameters and component properties in Table \ref{tab:81param} and
Figure \ref{fig:pl}.

In these eclipsing systems, the radial displacement of the Cepheid can
be detected directly solely from modelling of the photometric
lightcurve.  This is in contrast to most measurements of radial
amplitude which are inferred from radial velocity measurements.  As a
test, the fits of all three systems were repeated with the amplitude
of the radial displacement held fixed at 0 and the resulting $\chi^2$
values compared to those with radial variation included.  For the
system 6.6454.5, $\chi_v^2$ increased to 4.0, a 145\% increase over
the value of 1.6 listed in Table \ref{tab:6param}.  For 81.8997.87
$\chi_v^2$ increased to 1.5, an increase of only 26\% over the best
fit value of 1.1.  For 78.6338.24 $\chi^2$ increased to 8.8 from
7.5, only a 17\% change.  For each of the systems the change in
$\chi^2$ is found to be statistically significant for the number of
degrees of freedom present.  The relatively smaller impact of the
radial amplitude on the quality of the fit for 78.6338.24 could be
explained by at least two factors:
\begin{enumerate}

\item {In this system the eclipse duration and pulsation timescales
are sufficiently similar that they produce a degeneracy in a parameter
set which includes $\Delta$R}.

\item {The $\chi^2$ value for this system is already elevated due to
the model inadequacies previously mentioned.  Their impact on the fit
could easily dwarf the effects of the inclusion of radial amplitude.}

\end{enumerate}

For 81.8997.87 the small fractional amplitude of the radial change
($\sim$6\%) could make its effect on the lightcurve difficult to
discern.

\section{SUMMARY AND FUTURE WORK}

We have analysed the available multicolor photometry for three
eclipsing Cepheids. The principal results of this work are:

\begin{enumerate}

\item{MACHO Project V and R photometry for these important systems is reported.}

\item{The characterization of the three systems using a
self-consistent eclipsing binary lightcurve model which includes the
effects of radius and surface brightness change due to the pulsating
star.}

\item{The interpretation of the evolutionary state of the three
systems, with two of the three systems conforming poorly to the
expectations from standard, single star evolutionary theory.}

\item{Direct evidence of radial size change in a Cepheid variable. To
date, this has been implied (apparently correctly!) by integrating the
radial velocity curves of Cepheids. Michelson interferometry is
capable of measuring the angular diameters of Cepheids but to date the
change in angular diameter, which is direct evidence for radius
change, has only been reported for a single Cepheid \citep{lane}.}

\item{The prediction of future primary and secondary eclipses and
their fine structure to facilitate follow-up observations (see
below).}

\item{The identification of the factors which limit the usefulness of
the lightcurve model which has been developed.}

\end{enumerate}

Follow-up photometric observations are clearly warranted for all of
these systems, but especially for 81.8997.87 which is clearly an
intermediate-mass, first-overtone Cepheid. Indeed, several of us have
obtained multicolor photometry of the April 2001 primary minimum of
this system using the 74-inch telescope at Mount Stromlo Observatory,
Canberra, Australia. More densely spaced observations of this system
during a future secondary minimum, especially longer wavelength
photometry, are likely to provide significantly improved values for
surface brightness ratios.  To facilitate future observations
predicted dates of primary and secondary minima are listed for all
three systems in Tables \ref{tab:future6}-\ref{tab:future81}.

The full impact of the discovery of eclipsing Cepheids in the LMC will
only be realized when radial velocity curves for both components in
each system have been obtained. Recently, Cycle 10 HST (Hubble Space
Telecope) observations were taken by a collaboration involving several
coauthors on this paper and led by Edward Guinan of Villanova
University. For 81.8997.87, it appears that spectral observations in
the far red will be necessary to improve the contrast of the secondary
stars spectral lines relative to the Cepheid.  The long orbital
periods of these systems imply numerous HST visits over the course of
approximately a year to map out the radial velocity curves
completely. Fortunately, the relatively small amplitude of the radial
velocity curve of the Cepheid will result in a clean separation of the
orbital and pulsation components with a small number of visits.

We are very grateful for the skilled support given our project by the
technical staff at the Mount Stromlo Observatory, and in particular we
would like to thank Simon Chan, Glen Thorpe, Susannah Sabine, and
Michael McDonald for their invaluable assistance in obtaining the
data.  This work was performed under the auspices of the
U.S. Department of Energy by the University of California, Lawrence
Livermore National Laboratory under contract No. W-7405-Eng-48. Work
performed by the Center for Particle Astrophysics personnel is
supported in part by the Office of Science and Technology Centers of
the NSF under cooperative agreement A-8809616. Work performed at MSSSO
is supported by the Bilateral Science and Technology Program of the
Australian Department of Industry, Technology and Regional
Development.  DM is supported by FONDAP.  D. L. W. and D. Lepischak
were supported in part by a Research Grant from the Natural Sciences
and Engineering Research Council of Canada (NSERC) during this work.
This work comprised part of the M.Sc. thesis of D. Lepischak.


\begin{deluxetable}{cccc}
\small
\tablecolumns{4}
\tablewidth{0pc}
\tablecaption{V Photometry for 6.6454.5\label{tab-6.6454.5v}} 
\tablehead{
\colhead{HJD}&
\colhead{V}&
\colhead{$\sigma_V$}&
\colhead{Source}
\\
&\colhead{\sl (mag)}
&\colhead{\sl (mag)}
}
\startdata
{\tt2449074.95500}&{\tt14.6530}&{\tt0.0150}&{\tt MACHO}\\
{\tt2449075.06470}&{\tt14.6520}&{\tt0.0150}&{\tt MACHO}\\
{\tt2449075.96150}&{\tt14.6840}&{\tt0.0150}&{\tt MACHO}\\
{\tt2449076.99790}&{\tt14.7080}&{\tt0.0150}&{\tt MACHO}\\
{\tt2449077.11200}&{\tt14.7190}&{\tt0.0150}&{\tt MACHO}\\
{\tt2449081.02760}&{\tt14.7020}&{\tt0.0150}&{\tt MACHO}\\
{\tt2449082.99210}&{\tt14.6240}&{\tt0.0160}&{\tt MACHO}\\
{\tt2449083.94080}&{\tt14.6240}&{\tt0.0160}&{\tt MACHO}\\
{\tt2449084.97280}&{\tt14.6290}&{\tt0.0150}&{\tt MACHO}\\
{\tt2449085.94290}&{\tt14.6740}&{\tt0.0160}&{\tt MACHO}\\
{\tt2449088.06740}&{\tt14.6020}&{\tt0.0150}&{\tt MACHO}\\
{\tt2449088.96770}&{\tt14.6030}&{\tt0.0150}&{\tt MACHO}\\
{\tt2449089.97160}&{\tt14.6390}&{\tt0.0150}&{\tt MACHO}\\
{\tt2449093.90660}&{\tt14.5990}&{\tt0.0150}&{\tt MACHO}\\
{\tt2449098.04000}&{\tt14.6050}&{\tt0.0150}&{\tt MACHO}\\
\enddata
\end{deluxetable}

\begin{deluxetable}{cccc}
\small
\tablecolumns{4}
\tablewidth{0pc}
\tablecaption{R Photometry for 6.6454.5\label{tab-6.6454.5r}} 
\tablehead{
\colhead{HJD}&
\colhead{R}&
\colhead{$\sigma_R$}&
\colhead{Source}
\\
&\colhead{\sl (mag)}
&\colhead{\sl (mag)}
}
\startdata
{\tt2448824.17260}&{\tt14.5560}&{\tt0.0160}&{\tt MACHO}\\
{\tt2448827.18660}&{\tt14.5590}&{\tt0.0150}&{\tt MACHO}\\
{\tt2448829.14130}&{\tt14.5420}&{\tt0.0160}&{\tt MACHO}\\
{\tt2448830.15910}&{\tt14.5260}&{\tt0.0150}&{\tt MACHO}\\
{\tt2448833.15680}&{\tt14.6010}&{\tt0.0150}&{\tt MACHO}\\
{\tt2448834.27580}&{\tt14.5370}&{\tt0.0150}&{\tt MACHO}\\
{\tt2448835.14270}&{\tt14.4810}&{\tt0.0150}&{\tt MACHO}\\
{\tt2448836.21040}&{\tt14.5340}&{\tt0.0150}&{\tt MACHO}\\
{\tt2448837.13050}&{\tt14.5520}&{\tt0.0150}&{\tt MACHO}\\
{\tt2448843.16980}&{\tt14.6000}&{\tt0.0160}&{\tt MACHO}\\
{\tt2448844.13490}&{\tt14.5380}&{\tt0.0150}&{\tt MACHO}\\
{\tt2448851.15840}&{\tt14.5840}&{\tt0.0150}&{\tt MACHO}\\
{\tt2448854.28290}&{\tt14.5080}&{\tt0.0160}&{\tt MACHO}\\
{\tt2448855.25190}&{\tt14.5290}&{\tt0.0150}&{\tt MACHO}\\
{\tt2448857.30420}&{\tt14.6080}&{\tt0.0150}&{\tt MACHO}\\
\enddata
\end{deluxetable}

\begin{deluxetable}{cccc}
\small
\tablecolumns{4}
\tablewidth{0pc}
\tablecaption{I Photometry for 6.6454.5\label{tab-6.6454.5i}} 
\tablehead{
\colhead{HJD}&
\colhead{I}&
\colhead{$\sigma_I$}&
\colhead{Source}
\\
&\colhead{\sl (mag)}
&\colhead{\sl (mag)}
}
\startdata
{\tt2450832.78089}&{\tt14.4520}&{\tt0.0080}&{\tt OGLE}\\
{\tt2450834.74339}&{\tt14.3600}&{\tt0.0150}&{\tt OGLE}\\
{\tt2450836.68017}&{\tt14.4310}&{\tt0.0140}&{\tt OGLE}\\
{\tt2450838.75705}&{\tt14.3800}&{\tt0.0100}&{\tt OGLE}\\
{\tt2450839.71152}&{\tt14.3690}&{\tt0.0100}&{\tt OGLE}\\
{\tt2450840.69671}&{\tt14.3890}&{\tt0.0070}&{\tt OGLE}\\
{\tt2450841.77146}&{\tt14.4440}&{\tt0.0060}&{\tt OGLE}\\
{\tt2450842.69221}&{\tt14.4470}&{\tt0.0110}&{\tt OGLE}\\
{\tt2450843.69775}&{\tt14.3800}&{\tt0.0100}&{\tt OGLE}\\
{\tt2450844.69179}&{\tt14.3490}&{\tt0.0070}&{\tt OGLE}\\
{\tt2450845.55570}&{\tt14.3901}&{\tt0.0079}&{\tt CTIO}\\
{\tt2450845.80135}&{\tt14.3850}&{\tt0.0080}&{\tt OGLE}\\
{\tt2450846.75718}&{\tt14.4380}&{\tt0.0070}&{\tt OGLE}\\
{\tt2450850.73583}&{\tt14.3890}&{\tt0.0110}&{\tt OGLE}\\
{\tt2450851.65660}&{\tt14.4245}&{\tt0.0079}&{\tt CTIO}\\
\enddata

\end{deluxetable}
\begin{deluxetable}{cccc}
\small
\tablecolumns{4}
\tablewidth{0pc}
\tablecaption{V Photometry for 78.6338.24\label{tab-78.6338.24}} 
\tablehead{
\colhead{HJD}&
\colhead{V}&
\colhead{$\sigma_V$}&
\colhead{Source}
\\
&\colhead{\sl (mag)}
&\colhead{\sl (mag)}
}
\startdata
{\tt2448886.25480}&{\tt15.5390}&{\tt0.0150}&{\tt MACHO}\\
{\tt2448888.15230}&{\tt15.6790}&{\tt0.0170}&{\tt MACHO}\\
{\tt2448889.20710}&{\tt15.6940}&{\tt0.0160}&{\tt MACHO}\\
{\tt2448893.23740}&{\tt15.8760}&{\tt0.0170}&{\tt MACHO}\\
{\tt2448895.21320}&{\tt16.0230}&{\tt0.0190}&{\tt MACHO}\\
{\tt2448896.18200}&{\tt16.1130}&{\tt0.0160}&{\tt MACHO}\\
{\tt2448897.24090}&{\tt16.0870}&{\tt0.0170}&{\tt MACHO}\\
{\tt2448902.27770}&{\tt15.5770}&{\tt0.0150}&{\tt MACHO}\\
{\tt2448917.16550}&{\tt15.8170}&{\tt0.0160}&{\tt MACHO}\\
{\tt2448919.19520}&{\tt15.4720}&{\tt0.0150}&{\tt MACHO}\\
{\tt2448924.20640}&{\tt15.6990}&{\tt0.0150}&{\tt MACHO}\\
{\tt2448933.25020}&{\tt16.0810}&{\tt0.0240}&{\tt MACHO}\\
{\tt2448940.15000}&{\tt15.6570}&{\tt0.0160}&{\tt MACHO}\\
{\tt2448942.17400}&{\tt15.7230}&{\tt0.0160}&{\tt MACHO}\\
{\tt2448949.13640}&{\tt16.1180}&{\tt0.0160}&{\tt MACHO}\\
\enddata

\end{deluxetable}

\begin{deluxetable}{cccc}
\small
\tablecolumns{4}
\tablewidth{0pc}
\tablecaption{R Photometry for 78.6338.24\label{tab-78.6338.24r}} 
\tablehead{
\colhead{HJD}&
\colhead{R}&
\colhead{$\sigma_R$}&
\colhead{Source}
\\
&\colhead{\sl (mag)}
&\colhead{\sl (mag)}
}
\startdata
{\tt2448886.25480}&{\tt15.2120}&{\tt0.0150}&{\tt MACHO}\\
{\tt2448888.15230}&{\tt15.3130}&{\tt0.0170}&{\tt MACHO}\\
{\tt2448889.20710}&{\tt15.3140}&{\tt0.0160}&{\tt MACHO}\\
{\tt2448893.23740}&{\tt15.4970}&{\tt0.0170}&{\tt MACHO}\\
{\tt2448895.21320}&{\tt15.6400}&{\tt0.0180}&{\tt MACHO}\\
{\tt2448896.18200}&{\tt15.7300}&{\tt0.0150}&{\tt MACHO}\\
{\tt2448897.24090}&{\tt15.7460}&{\tt0.0170}&{\tt MACHO}\\
{\tt2448902.27770}&{\tt15.2600}&{\tt0.0150}&{\tt MACHO}\\
{\tt2448917.16550}&{\tt15.5060}&{\tt0.0160}&{\tt MACHO}\\
{\tt2448919.19520}&{\tt15.1810}&{\tt0.0150}&{\tt MACHO}\\
{\tt2448924.20640}&{\tt15.3400}&{\tt0.0150}&{\tt MACHO}\\
{\tt2448933.25020}&{\tt15.7500}&{\tt0.0270}&{\tt MACHO}\\
{\tt2448940.15000}&{\tt15.2980}&{\tt0.0160}&{\tt MACHO}\\
{\tt2448942.17400}&{\tt15.3340}&{\tt0.0160}&{\tt MACHO}\\
{\tt2448949.13640}&{\tt15.7660}&{\tt0.0150}&{\tt MACHO}\\
\enddata

\end{deluxetable}

\begin{deluxetable}{cccc}
\small
\tablecolumns{4}
\tablewidth{0pc}
\tablecaption{V Photometry for 81.8997.87\label{tab-81.8997.87v}} 
\tablehead{
\colhead{HJD}&
\colhead{V}&
\colhead{$\sigma_V$}&
\colhead{Source}
\\
&\colhead{\sl (mag)}
&\colhead{\sl (mag)}
}
\startdata
{\tt2448919.24260}&{\tt17.1530}&{\tt0.0220}&{\tt MACHO}\\
{\tt2448920.04760}&{\tt17.0300}&{\tt0.0230}&{\tt MACHO}\\
{\tt2448921.04080}&{\tt17.1880}&{\tt0.0220}&{\tt MACHO}\\
{\tt2448924.25670}&{\tt16.9930}&{\tt0.0230}&{\tt MACHO}\\
{\tt2448927.99640}&{\tt16.9700}&{\tt0.0270}&{\tt MACHO}\\
{\tt2448929.03000}&{\tt17.1960}&{\tt0.0200}&{\tt MACHO}\\
{\tt2448929.97740}&{\tt16.9830}&{\tt0.0240}&{\tt MACHO}\\
{\tt2448930.97620}&{\tt17.1670}&{\tt0.0360}&{\tt MACHO}\\
{\tt2448931.99100}&{\tt17.0060}&{\tt0.0240}&{\tt MACHO}\\
{\tt2448932.16440}&{\tt16.9770}&{\tt0.0240}&{\tt MACHO}\\
{\tt2448934.13420}&{\tt16.9560}&{\tt0.0220}&{\tt MACHO}\\
{\tt2448936.13460}&{\tt16.9940}&{\tt0.0420}&{\tt MACHO}\\
{\tt2448938.18980}&{\tt16.9860}&{\tt0.0300}&{\tt MACHO}\\
{\tt2448939.01450}&{\tt17.1340}&{\tt0.0280}&{\tt MACHO}\\
{\tt2448940.03150}&{\tt17.0110}&{\tt0.0240}&{\tt MACHO}\\
\enddata

\end{deluxetable}

\begin{deluxetable}{cccc}
\small
\tablecolumns{4}
\tablewidth{0pc}
\tablecaption{R Photometry for 81.8997.87\label{tab-81.8997.87r}} 
\tablehead{
\colhead{HJD}&
\colhead{R}&
\colhead{$\sigma_R$}&
\colhead{Source}
\\
&\colhead{\sl (mag)}
&\colhead{\sl (mag)}
}
\startdata
{\tt2448919.24260}&{\tt16.4420}&{\tt0.0170}&{\tt MACHO}\\
{\tt2448920.04760}&{\tt16.3470}&{\tt0.0180}&{\tt MACHO}\\
{\tt2448921.04080}&{\tt16.4710}&{\tt0.0170}&{\tt MACHO}\\
{\tt2448924.25670}&{\tt16.3450}&{\tt0.0180}&{\tt MACHO}\\
{\tt2448927.99640}&{\tt16.3040}&{\tt0.0190}&{\tt MACHO}\\
{\tt2448929.03000}&{\tt16.4600}&{\tt0.0170}&{\tt MACHO}\\
{\tt2448929.97740}&{\tt16.3200}&{\tt0.0180}&{\tt MACHO}\\
{\tt2448930.97620}&{\tt16.4390}&{\tt0.0210}&{\tt MACHO}\\
{\tt2448931.99100}&{\tt16.3390}&{\tt0.0180}&{\tt MACHO}\\
{\tt2448932.16440}&{\tt16.3170}&{\tt0.0180}&{\tt MACHO}\\
{\tt2448934.13420}&{\tt16.2960}&{\tt0.0170}&{\tt MACHO}\\
{\tt2448936.13460}&{\tt16.3140}&{\tt0.0240}&{\tt MACHO}\\
{\tt2448938.18980}&{\tt16.3100}&{\tt0.0210}&{\tt MACHO}\\
{\tt2448939.01450}&{\tt16.4250}&{\tt0.0190}&{\tt MACHO}\\
{\tt2448940.03150}&{\tt16.3440}&{\tt0.0180}&{\tt MACHO}\\
\enddata

\end{deluxetable}

\begin{deluxetable}{cccc}
\small
\tablecolumns{4}
\tablewidth{0pc}
\tablecaption{I Photometry for 81.8997.87\label{tab-81.8997.87i}} 
\tablehead{
\colhead{HJD}&
\colhead{I}&
\colhead{$\sigma_I$}&
\colhead{Source}
\\
&\colhead{\sl (mag)}
&\colhead{\sl (mag)}
}
\startdata
{\tt2450739.86067}&{\tt15.6720}&{\tt0.0100}&{\tt OGLE}\\
{\tt2450744.76901}&{\tt15.7840}&{\tt0.0130}&{\tt OGLE}\\
{\tt2450745.85362}&{\tt15.6680}&{\tt0.0130}&{\tt OGLE}\\
{\tt2450746.83950}&{\tt15.7950}&{\tt0.0180}&{\tt OGLE}\\
{\tt2450747.73495}&{\tt15.6580}&{\tt0.0130}&{\tt OGLE}\\
{\tt2450750.79988}&{\tt15.8020}&{\tt0.0130}&{\tt OGLE}\\
{\tt2450751.80069}&{\tt15.6610}&{\tt0.0200}&{\tt OGLE}\\
{\tt2450752.85852}&{\tt15.7830}&{\tt0.0110}&{\tt OGLE}\\
{\tt2450755.78274}&{\tt15.6530}&{\tt0.0100}&{\tt OGLE}\\
{\tt2450759.70861}&{\tt15.6680}&{\tt0.0150}&{\tt OGLE}\\
{\tt2450761.78982}&{\tt15.6860}&{\tt0.0130}&{\tt OGLE}\\
{\tt2450766.83954}&{\tt15.7490}&{\tt0.0150}&{\tt OGLE}\\
{\tt2450773.79513}&{\tt15.7050}&{\tt0.0140}&{\tt OGLE}\\
{\tt2450776.82033}&{\tt15.7410}&{\tt0.0140}&{\tt OGLE}\\
{\tt2450778.85342}&{\tt15.7100}&{\tt0.0100}&{\tt OGLE}\\
\enddata

\end{deluxetable}

\clearpage

\begin{deluxetable}{llll}
\small
\tablecolumns{4}
\tablewidth{0pc}
\tablecaption{Best fit parameters for 6.6454.5.  Meanings of individual parameters and units are explained in the text.  \label{tab:6param}}
\tableheadfrac{0}
\tablehead{
\multicolumn{1}{l}{$\chi_v^2$} & \multicolumn{1}{l}{1.6}&&
}
\startdata
$P_{orb}~(d)$&397.142 $\pm$ 0.005&&\\ 
$i~(^\circ)$&86.71 $\pm$ 0.04&&\\
\tableline
Companion&(primary)&Variable&(secondary)\\ 
\tableline
$R$&0.0522 $\pm$ 0.0004	&$R_{min}$&0.051 $\pm$ 0.001\\ 
$\frac{J_V}{J_R}$&1.061 $\pm$ 0.002&$<\frac{J_V}{J_R}>$&0.748 $\pm$ 0.005\\ 
$x_V$&0.46 $\pm$ 0.17&$x_V$&0.52 $\pm$ 0.85\\ 
$x_R$&0.48 $\pm$ 0.20&$x_R$&0.42 $\pm$ 0.72\\ 
$x_I$&0.45 $\pm$ 0.18&$x_I$&0.56 $\pm$ 0.55\\ 
&&$P_{ceph}~(d)$&4.97371 $\pm$ 0.00002\\ 
&&$\Delta R_{amp}$&0.0166 $\pm$ 0.0002\\ 
&&$\Delta R_{shift}~(d)$&-0.17 $\pm$ 0.01\\ 

\cutinhead{Intensity-weighted Mean Magnitudes and Colours}
$V$&14.82 $\pm$ 0.04&$<V>$& 16.7$\pm$0.1 \\ 
$R$&14.80 $\pm$ 0.04&$<R>$& 16.3$\pm$ 0.1\\ 
$I$&14.72 $\pm$ 0.04&$<I>$& 15.89$\pm$ 0.09\\ 
$V-R$&0.018 $\pm$ 0.002&$<V-R>$&0.452 $\pm$ 0.007\\ 
$V-I$& 0.102$\pm$ 0.003&$<V-I>$& 0.81$\pm$ 0.01\\ 
&&$<W_R>$& 14.4$\pm$0.1 \\ 


\enddata
\end{deluxetable}

\begin{deluxetable}{llll}
\small
\tablecolumns{4}
\tablewidth{0pc}
\tablecaption{Best fit parameters for 78.6338.24.  Parameter definitions and units are the same as in Table \ref{tab:6param} with the exception of $A_1$ and $A_2$ which are defined by equation \ref{eq:freq}.\label{tab:78param}}
\tableheadfrac{0}
\tablehead{
\multicolumn{1}{l}{$\chi_v^2$} & \multicolumn{1}{l}{7.5}&&
}
\startdata
$P_{orb}~(d)$&419.718 $\pm$ 0.008&&\\ 
$i~(^\circ)$&86.94 $\pm$ 0.02&&\\
\tableline
Companion&(primary)&Variable&(secondary)\\ 
\tableline
$R$&0.0403 $\pm$ 0.0002 &$R_{min}$&0.0717 $\pm$ 0.0004\\ 
$\frac{J_V}{J_R}$&1.146 $\pm$ 0.001&$<\frac{J_V}{J_R}>$&0.9118 $\pm$ 0.001\\ 
$x_V$&0.46 $\pm$ 0.17&$x_V$&0.52 $\pm$ 0.85\\ 
$x_R$&0.48 $\pm$ 0.20&$x_R$&0.42 $\pm$ 0.72\\ 
&&$P_{ceph}~(d)$&17.68586 $\pm$ 0.0003\\ 
&&$A_1$&2.265 $\pm$ 0.007\\ 
&&$A_2~(d)$&-448 $\pm$ 2\\ 
&&$\Delta R_{amp}$&0.03106 $\pm$ 0.0002\\ 
&&$\Delta R_{shift}~(d)$&-6.07 $\pm$ 0.01\\ 

\cutinhead{Intensity-weighted Mean Magnitudes and Colours}
$V$&16.55 $\pm$ 0.02&$<V>$& 16.20$\pm$0.03 \\ 
$R$&16.35 $\pm$ 0.02&$<R>$& 15.75$\pm$ 0.03\\ 
$V-R$&0.20 $\pm$ 0.03&$<V-R>$&0.447 $\pm$ 0.001\\ 
&&$<W_R>$& 13.98$\pm$0.03 \\ 


\enddata
\end{deluxetable}

\begin{deluxetable}{llll}
\small \tablecolumns{4} \tablewidth{0pc} \tablecaption{Best-fit
parameters for 81.8997.87.  Parameters are the same as in Table
\ref{tab:6param}.  The large uncertainties in the limb-darkening
coefficients of the secondary indicate that while these parameters are
allowed to vary, they are essentially unmodified and unconstrained by
the lightcurve.  The tabulated $x_\lambda$ are the original values and
the errors are included for consistency.\label{tab:81param}}
\tableheadfrac{0} \tablehead{
\multicolumn{1}{l}{$\chi_v^2$} & \multicolumn{1}{l}{1.2}&&
}
\startdata
$P_{orb}~(d)$&800.5 $\pm$ 0.1&&\\ 
$i~(^\circ)$&87.0 $\pm$ 0.2&&\\
\tableline
Variable&(primary)&Companion&(secondary)\\ 
\tableline
$R_{min}$&0.029$\pm$ 0.001	&$R$&0.047 $\pm$ 0.005\\ 
$<\frac{J_V}{J_R}>$&1.078 $\pm$ 0.009&$\frac{J_V}{J_R}$&0.54 $\pm$ 0.05\\ 
$x_V$&0.50 $\pm$ 0.94&$x_V$&0.50 $\pm$ 6.06\\ 
$x_R$&0.50 $\pm$ 0.91&$x_R$&0.50 $\pm$ 4.52\\ 
$x_I$&0.50 $\pm$ 0.98&$x_I$&0.50 $\pm$ 3.02\\ 
$P_{ceph}~(d)$&2.035321 $\pm$ 0.000009\\ 
$\Delta R_{amp}$&0.0017 $\pm$ 0.0002\\ 
$\Delta R_{shift}~(d)$&0.21 $\pm$ 0.01\\ 

\cutinhead{Intensity-weighted Mean Magnitudes and Colours}
$<V>$&17.2 $\pm$ 0.2&$V$& 20$\pm$1 \\ 
$<R>$&16.6 $\pm$ 0.2&$R$& 19.2$\pm$ 0.9\\ 
$<I>$&16.0 $\pm$ 0.2&$I$& 17.9$\pm$ 0.8\\ 
$<V-R>$&0.602 $\pm$ 0.009&$V-R$& 1.3$\pm$ 0.1\\ 
$<V-I>$& 1.137$\pm$ 0.007&$V-I$& 2.6$\pm$ 0.2\\ 
$<W_R>$& 14.1$\pm$0.2 \\ 


\enddata
\end{deluxetable}


\begin{deluxetable}{ll|ll}
\small 
\tablecolumns{4} 
\tablewidth{0pc} 
\tablecaption{Predicted dates
of future eclipses for 6.6454.5.\label{tab:future6}} 
\tablehead{
\multicolumn{2}{c}{Primary Eclipse} & \multicolumn{2}{c}{Secondary
Eclipse}\\ 
\multicolumn{1}{c}{JD} & \multicolumn{1}{c}{UT} &
\multicolumn{1}{c}{JD} & \multicolumn{1}{c}{UT} } 
\startdata
2452454.78 & 2002 Jun 28 18.79 & 2452653.37 & 2003 Jan 13 8.80\\
2452851.95 & 2003 Jul 30 22.80 & 2453050.53 & 2004 Feb 14 12.80\\
2453249.12 & 2004 Aug 31 2.80 & 2453447.70 & 2005 Mar 17 16.81\\
2453646.28 & 2005 Oct 2 6.81 & 2453844.87 & 2006 Apr 18 20.81\\
2454043.45 & 2006 Nov 3 10.81 & 2454242.03 & 2007 May 21 0.82\\
2454440.62 & 2007 Dec 5 14.82 & 2454639.20 & 2008 Jun 21 4.82\\
2454837.78 & 2009 Jan 5 18.82 & 2455036.37 & 2009 Jul 23 8.83\\
2455234.95 & 2010 Feb 6 22.83 & 2455433.53 & 2010 Aug 24 12.83\\
2455632.12 & 2011 Mar 11 2.83 & 2455830.70 & 2011 Sep 25 16.84\\
\enddata
\end{deluxetable}

\begin{deluxetable}{ll|ll}
\small
\tablecolumns{4}
\tablewidth{0pc}
\tablecaption{Predicted dates of future eclipses for 78.6338.24.\label{tab:future78}}
\tablehead{
\multicolumn{2}{c}{Primary Eclipse} & \multicolumn{2}{c}{Secondary Eclipse}\\
\multicolumn{1}{c}{JD} & \multicolumn{1}{c}{UT} & \multicolumn{1}{c}{JD} & \multicolumn{1}{c}{UT}
}
\startdata
2452379.74 & 2002 Apr 14 17.70 & 2452589.47 & 2002 Nov 10 11.29\\
2452799.20 & 2003 Jun 8 4.88   & 2453008.94 & 2004 Jan 3 22.48\\
2453218.67 & 2004 Jul 31 16.07 & 2453428.40 & 2005 Feb 26 9.66\\
2453638.14 & 2005 Sep 24 3.25  & 2453847.87 & 2006 Apr 21 20.84\\
2454057.60 & 2006 Nov 17 14.43 & 2454267.33 & 2007 Jun 15 8.02\\
2454477.07 & 2008 Jan 11 1.61  & 2454686.80 & 2008 Aug 7 19.20\\
2454896.53 & 2009 Mar 5 12.79  & 2455106.27 & 2009 Oct 1 6.39\\
2455316.00 & 2010 Apr 28 23.98 & 2455525.73 & 2010 Nov 24 17.57\\
2455735.46 & 2011 Jun 22 11.16 & 2455945.20 & 2012 Jan 18 4.75\\
\enddata
\end{deluxetable}

\begin{deluxetable}{ll|ll}
\small
\tablecolumns{4}
\tablewidth{0pc}
\tablecaption{Predicted dates of future eclipses for 81.8997.87.\label{tab:future81}}
\tablehead{
\multicolumn{2}{c}{Primary Eclipse} & \multicolumn{2}{c}{Secondary Eclipse}\\
\multicolumn{1}{c}{JD} & \multicolumn{1}{c}{UT} & \multicolumn{1}{c}{JD} & \multicolumn{1}{c}{UT}
}
\startdata
2452010.81 & 2001 Apr 10 19.47 & 2452411.27 & 2002 May 16 6.45\\
2452811.73 & 2003 Jun 20 17.44 & 2453212.18 & 2004 Jul 25 4.42\\
2453612.64 & 2005 Aug 29 15.40 & 2454013.10 & 2006 Oct 4 2.38\\
2454413.56 & 2007 Nov 8 13.37  & 2454814.01 & 2008 Dec 13 0.35\\
2455214.47 & 2010 Jan 17 11.33 & 2455614.93 & 2011 Feb 21 22.32\\
2456015.39 & 2012 Mar 28 9.30  & 2456415.85 & 2013 May 2 20.28\\
2456816.30 & 2014 Jun 7 7.26   & 2457216.76 & 2015 Jul 12 18.25\\
2457617.22 & 2016 Aug 16 5.23  & 2458017.68 & 2017 Sep 20 16.21\\
2458418.13 & 2018 Oct 26 3.20  & 2458818.59 & 2019 Nov 30 14.18\\
2459219.05 & 2021 Jan 4 1.16   & 2459619.51 & 2022 Feb 8 12.14\\
\enddata
\end{deluxetable}

\begin{figure}
\epsscale{1.0}
\plotone{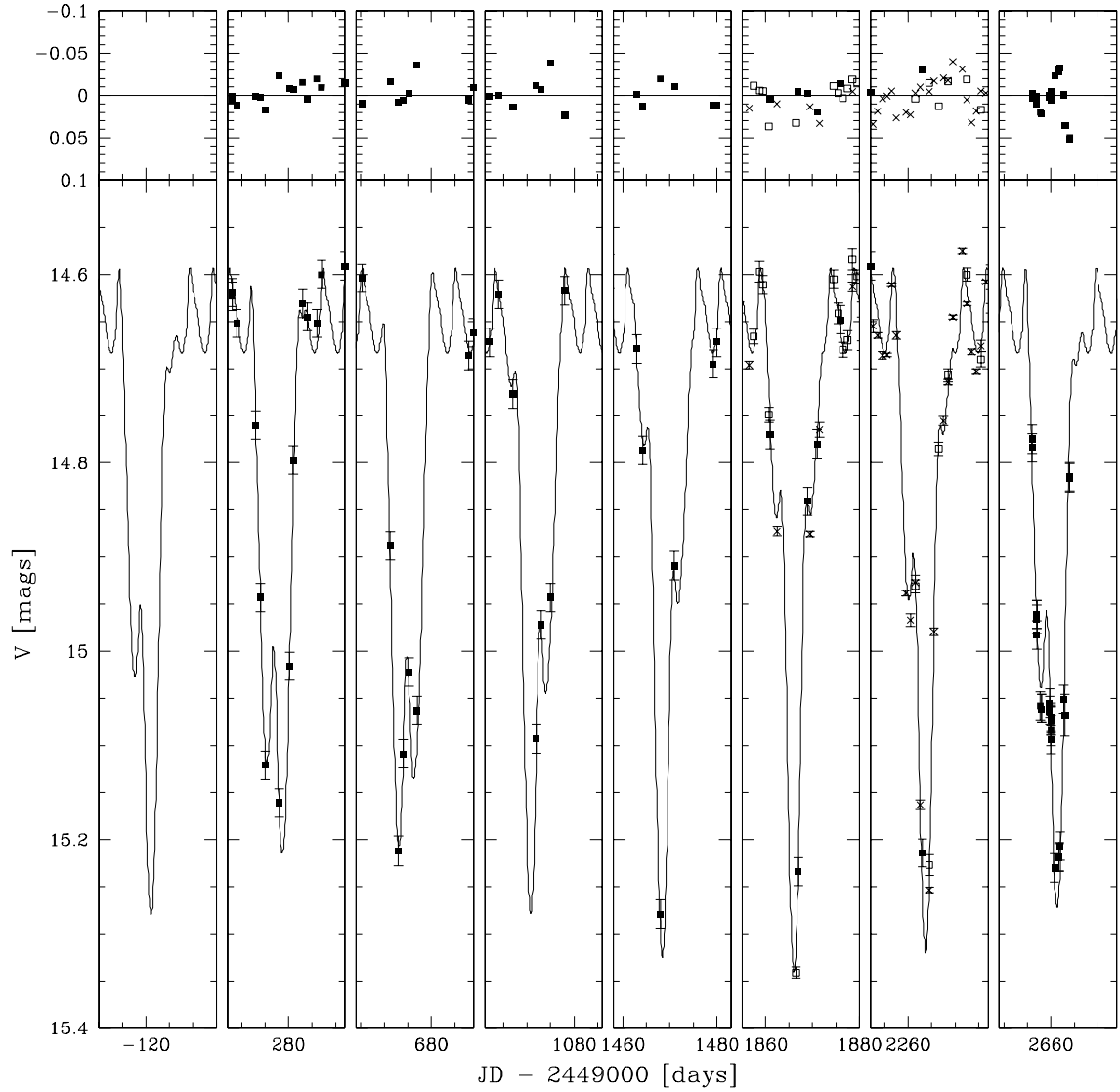}
\caption{Primary eclipses of 6.6454.4 in V with curve of best fit.
Upper panels show residuals in magnitudes.  Filled boxes indicate
observations from the MACHO project, open boxes indicate observations
from the OGLE project, x's are follow-up observations taken at
CTIO.\label{fig:6vp}}
\end{figure}

\begin{figure}
\epsscale{1.0}
\plotone{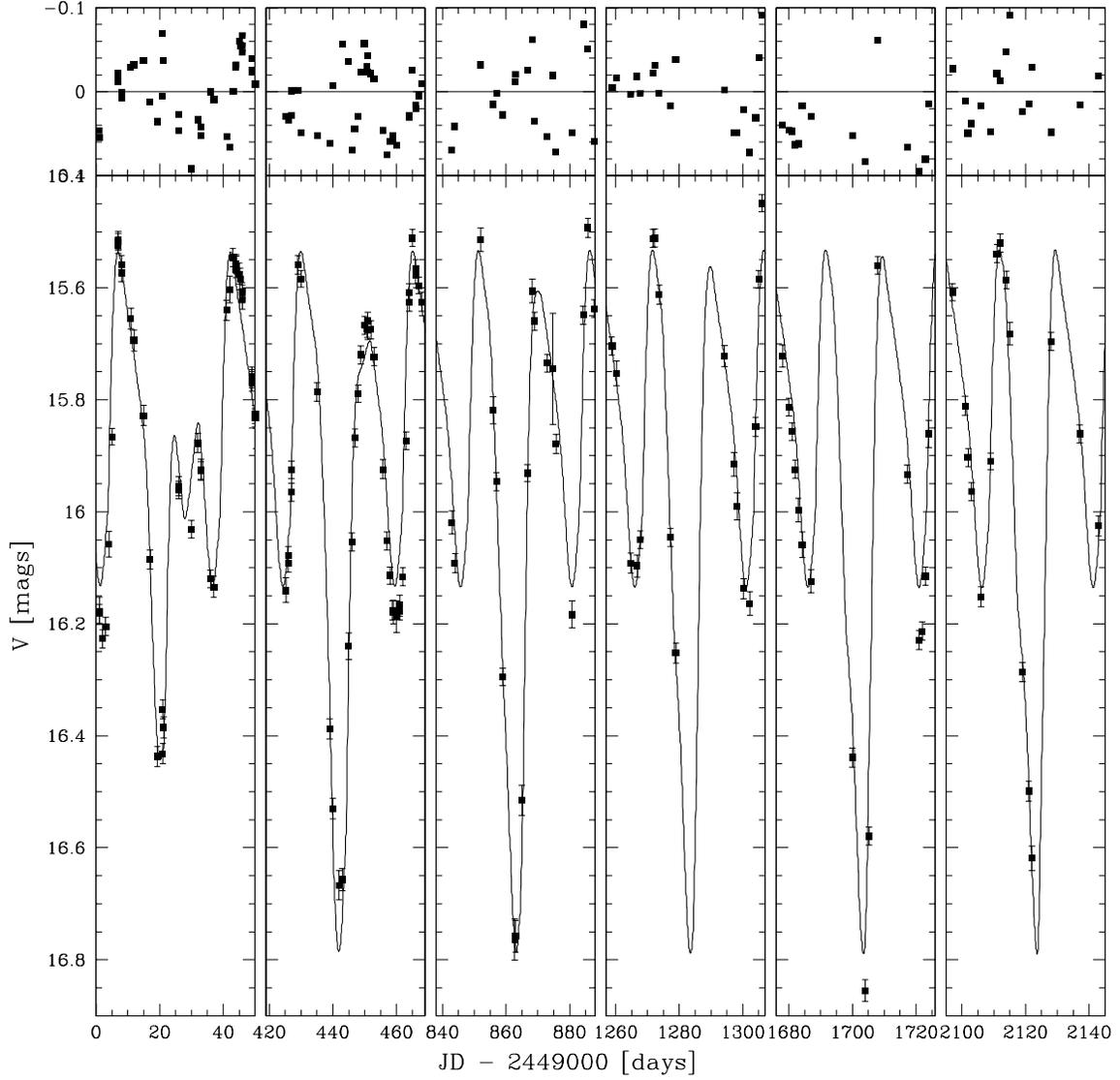}
\caption{Primary eclipses of 78.6338.24 in V with curve of best fit.
Upper panels show residuals in magnitudes. All points are from the
MACHO project database.\label{fig:78vp}}
\end{figure}

\begin{figure}
\epsscale{1.0}
\plotone{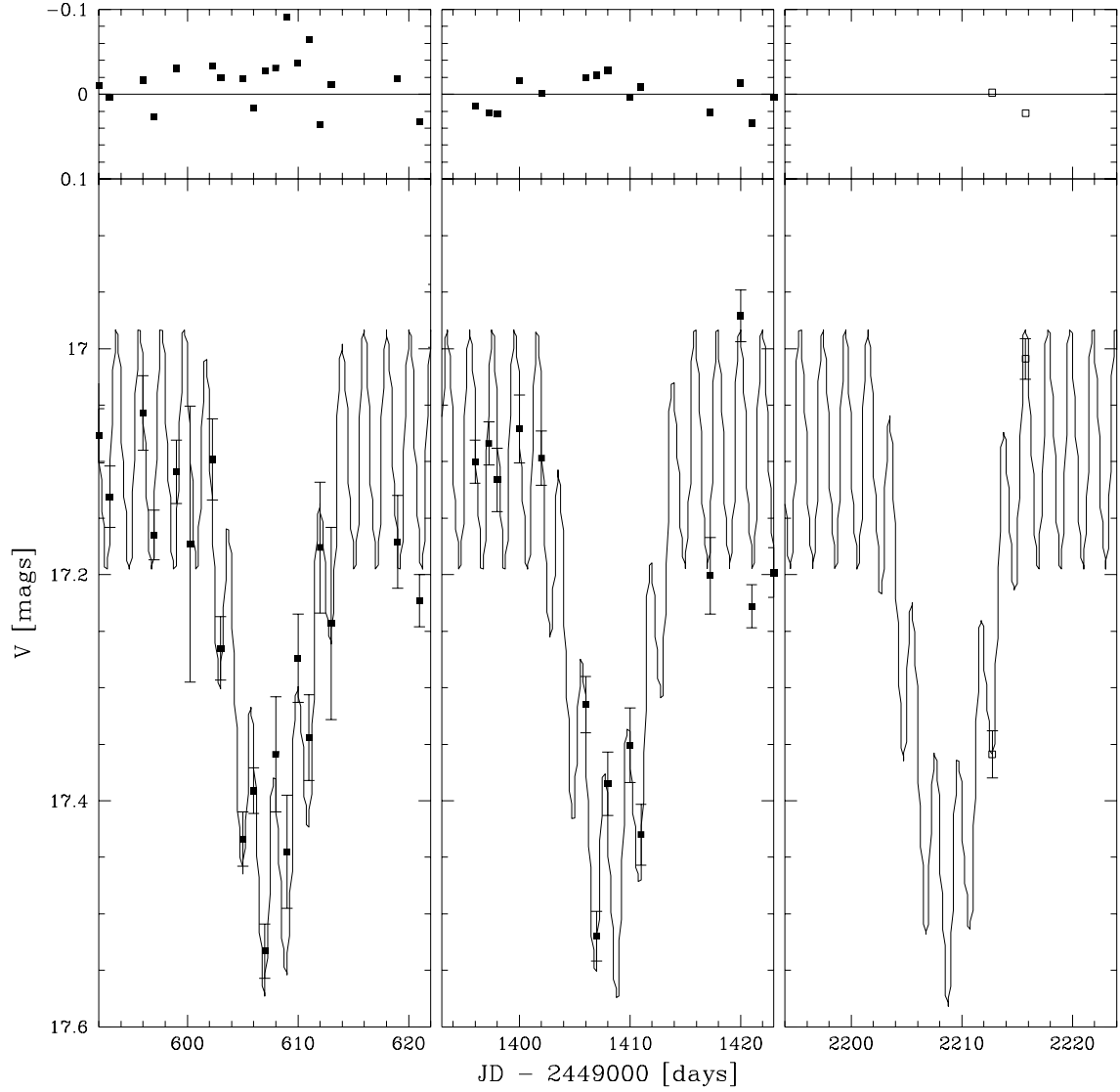}
\caption{Primary eclipses of 81.8997.87 in V with curve of best fit.
Upper panels show residuals in magnitudes.  Filled boxes indicate
observations from the MACHO project, open boxes observations from the
OGLE project. \label{fig:81vp}}
\end{figure}

\begin{figure}
\epsscale{1.0}
\plotone{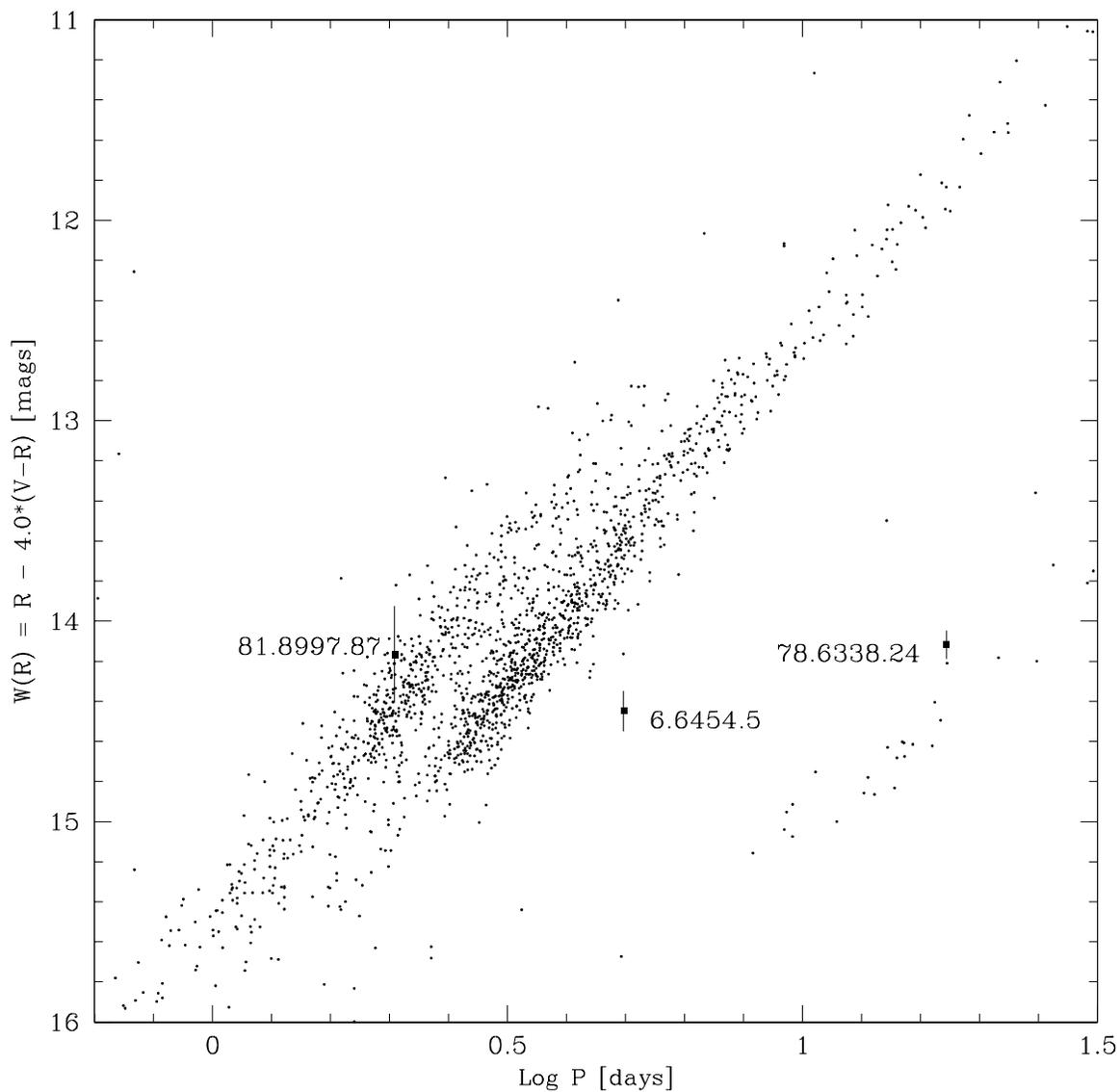}
\caption{$W_R$ vs. $log_{10} P$ diagram for 1766 MACHO Cepheids.  The
brighter sequence at a given period are Cepheids pulsating in the
first overtone and the sequence extending to longer periods are
fundamental mode pulsators.  Stars in the lower right are Type II
Cepheids.  The locations of the three Cepheids studied here are
indicated.  Their magnitudes and colors are from the best-fit
parameters and thus have had the companion flux removed.\label{fig:pl}}
\end{figure}

\begin{figure}
\epsscale{1.0}
\plotone{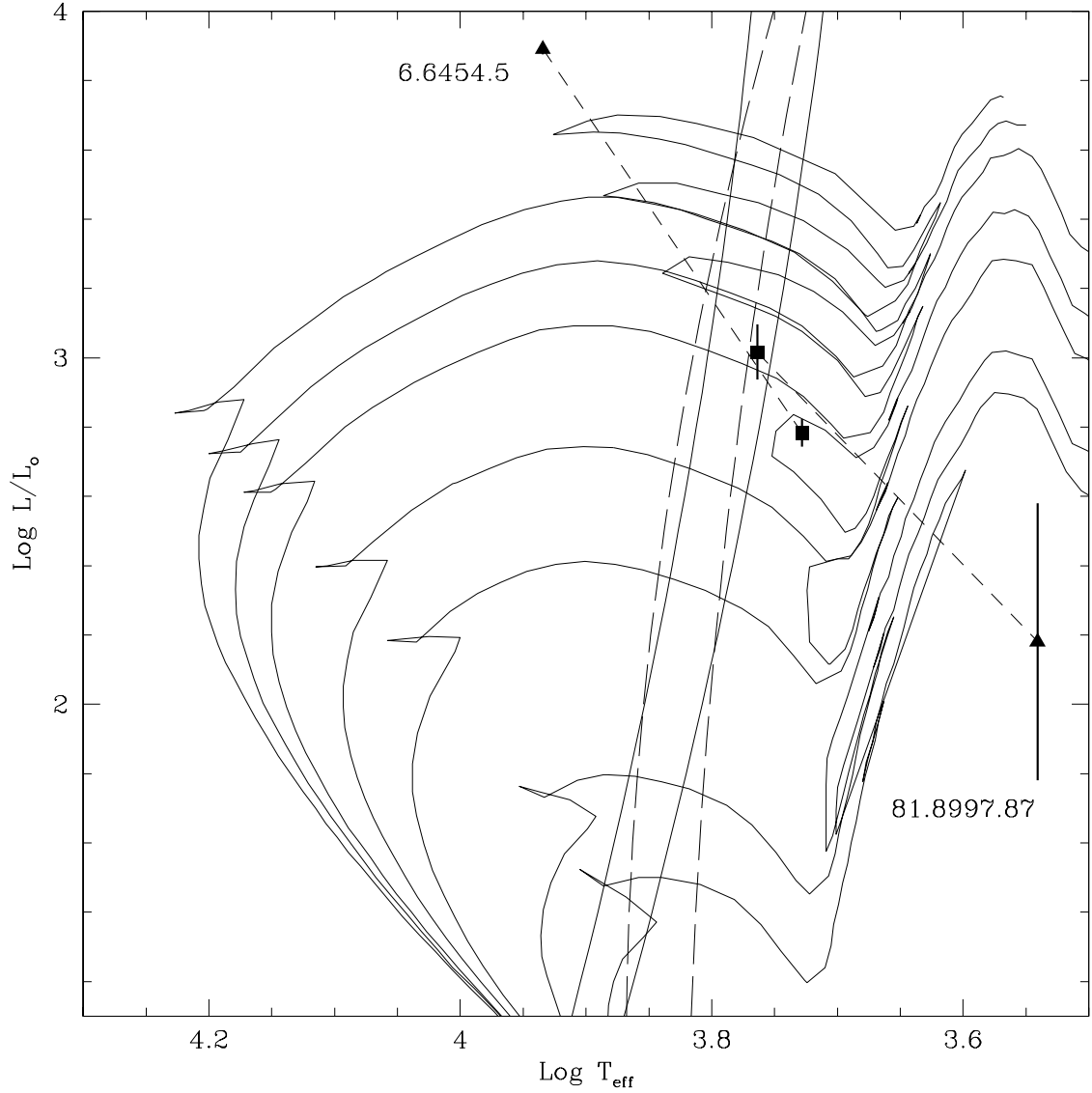}
\caption{Systems 6.6454.5 and 81.8997.87 compared with theoretical
isochrones for Y = 0.25, Z = 0.008 stars \citep{bertelli} ranging from
log(age)=9.1 to log(age)=7.9 [years].  Also shown are the theoretical
fundamental (solid) and overtone (dashed) instability strips of
\citet{cwc}.  Cepheids are shown as solid squares, their companions as
solid triangles.  Dashed lines connect system
members.\label{fig:iso1}}
\end{figure}

\begin{figure}
\epsscale{1.0}
\plotone{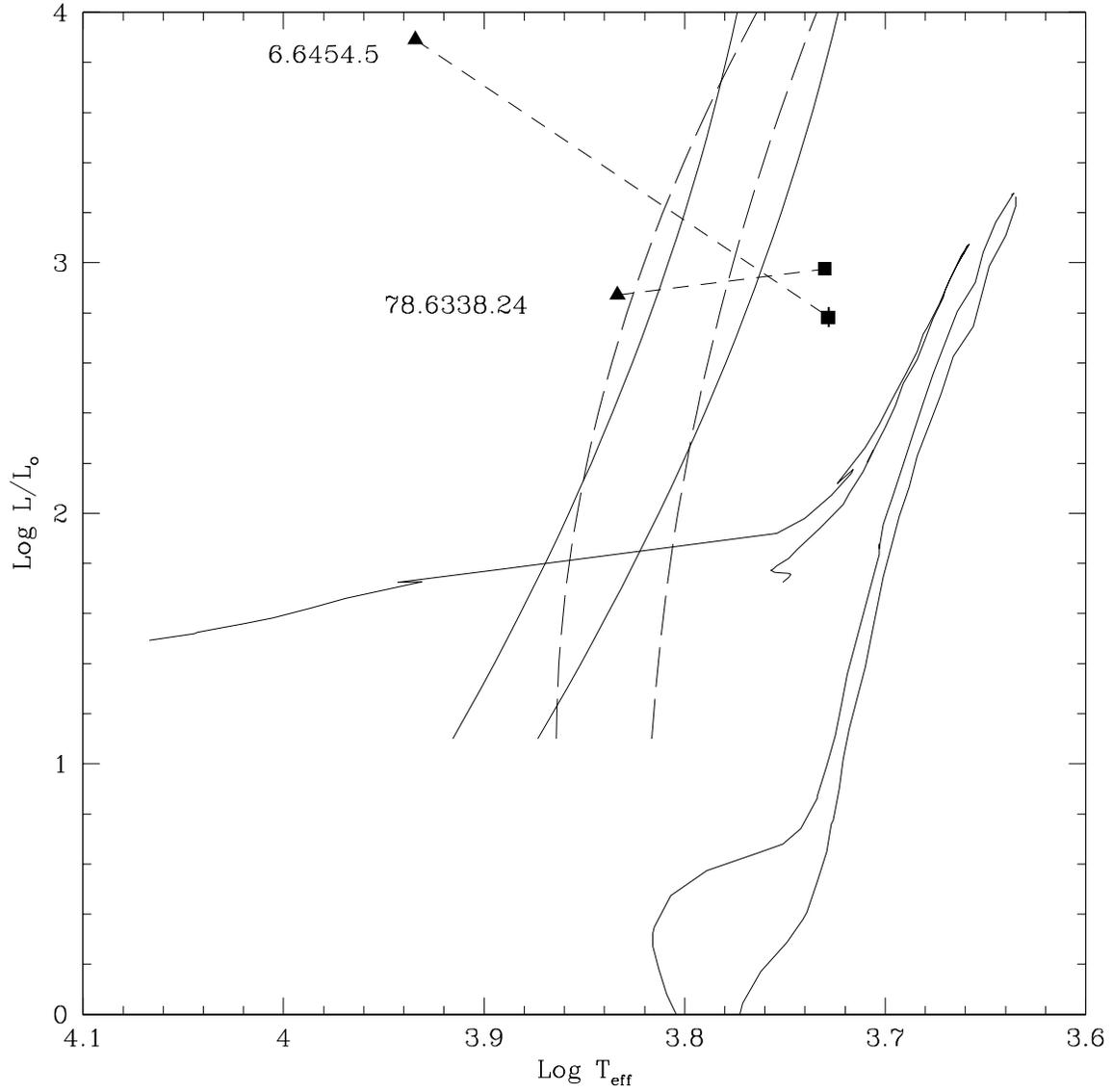}
\caption{Systems 6.6454.5 and 78.6338.24 compared with theoretical
isochrones for Y = 0.230, Z = 0.0004 stars \citep{fagotto}.  Also
shown are the theoretical fundamental (solid) and overtone (dashed)
instability strips of \citet{cwc}.\label{fig:iso2}}
\end{figure}

\end{document}